\documentstyle[12pt,twoside]{article}
\def\chapter{\clearpage\global\@topnum\z@
  \@afterindenttrue\secdef\@chapter\@schapter}  
\begin{document}
\pagestyle{plain}
\begin{flushright}
{\bf {\em DPNU-97-46}}\\
{\bf {\em BELLE-NOTE 190}}\\
\end{flushright}
{\ }\\
\begin{center}
\Large
{\ }\\
{\bf Examination of $T$/$CP$ Invariance \\
in the $e^+e^-\rightarrow\tau^+\tau^-$ Reaction} \\
\normalsize
{\ }\\
{\ }\\
{\bf T. Ohshima, S. Suito, A. Sugiyama, S. Suzuki} \\
{\em Nagoya University, Chikusa, Nagoya 464}\\
and \\
{\bf N. Haba} \\
{\em Mie University, Tsu, Mie 514} \\
{\ }\\
\scriptsize
\end{center}
\footnotesize
\begin{quote}
{\bf Abstract}\\
We propose a method to examine the $T/CP$ invariance of the heaviest 
lepton, $\tau$, by means of a triple-momentum correlation for 
the reaction $e^+e^-\rightarrow\tau^+\tau^-$, $\tau\rightarrow
e/\mu\nu\overline{\nu}$ during the course of the $B$-Factory 
experimental program. 
An unprecedented high sensitivity could be obtained 
without requiring a high integrated luminosity. \\
\end{quote}
{\ }\\
{\ }\\
\small
\noindent
{\bf 1. Introduction}\\
\par
Time reflection ($T$) is equivalent to the space-charge conjugation 
transformation ($CP$) under the $CPT$ invariant theorem.
To date, only the neutral kaon has exhibited the phenomenon of 
$CP$ violation at the ${\cal O}(10^{-3})$ level; the 
$B$-factory aims to scrutinize the violation mechanism, expecting 
its appearance in B-meson decay at the ${\cal O}(10^{-1})$ level 
through the so-called $CKM$ complex coupling in the Standard Model. 
On the other hand, $T$ invariance has been examined  
in many different processes, such as the electric-dipole moments of 
the neutron, electron, and atoms, neutron $\beta$-decay, triple correlations 
among initial- and final-state particles in nuclear-decays [1]. 
However, no search has been made for pure leptonic transitions of leptons. 
The only exception is a measurement of the triple correlation among 
the spin and momenta in $\mu\rightarrow e\nu\overline{\nu}$ decay, which 
found no violation with a sensitivity of 2.3\% [2]. \\

In a way, it is natural to expect the existence of $CP$ violation 
in the lepton sector as well as the quark sector; 
in particular, the leptonic decay of the heaviest lepton, $\tau$, 
could exhibit a larger violation than others, just like what happened 
in heavy quark transitions, $b\rightarrow u$ and $t\rightarrow d$. 
$CP$ violation in the lepton sector would appear beyond the Standard 
Model. 
For instance, the three-Higgs doublet model [4], and $R$-parity 
conserving [5] and violating [6] $SUSY$ models; also, Dirac or/and 
Majorana neutrino masses could provide one or more $CP$ phases. 
An examination of the $T$ and $CP$ invariances in any possible 
leptonic reactions should therefore be performed as accurately as 
experiments allow.  \\

T.D. Lee [3] recommended a study of $T$ and $CP$ violation in the 
$e^+e^-\rightarrow\tau^+\tau^-$ reaction, where the $\tau$'s decay 
pure-leptonically to both $(\mu/e)\nu\overline{\nu}$, in terms 
of a triple-momentum correlation as 
\begin{equation}
<{\cal A}> \equiv <~\hat{p_1}~\cdot~(\hat{p_2}\times\hat{p_3})~>,\
\end{equation}
where $\hat{p_1}$ is the unit vector of incident $e^-$ (or
$e^+$) momentum and $\hat{p_2}$ and $\hat{p_3}$ are the unit 
vectors of the outgoing $\mu$ and $e$ momenta, respectively. 
$<{\cal A}>$ means an average quantity comprised by these unit 
vectors, and yields a non-zero value 
if the $T/CP$ invariance does not hold. 
This triple correlation, $\cal A$, is odd under both $P$ and $T$
transformations, while the correlations in most experiments 
include the spin vector instead of the momentum vector of a respective 
particle; thus, $P$=even, but $T$=odd. 
Since the radiative correction 
would have 
an effect on the order of ${\cal O}(\alpha/2\pi)$, $T/CP$ violation 
can be simply established if $<{\cal A}>$ is larger [3]. \\

We consider here the experimental feasibility of testing the above 
mentioned lepton's $T/CP$ violation in the $e\mu$ final states of the 
$e^+e^-\rightarrow\tau^+\tau^-$ reaction by the $BELLE$ detector [7,8] 
at the $KEK-B$ Factory. 
Although, for the purpose of simplefying the arguments, 
we form a ratio ${\cal R}$, 
instead of $<{\cal A}>$, between the numbers of data samples with 
positive and negative ${\cal A}$ values, all of the following 
issues are in essence also valid for $<{\cal A}>$. 
In order to control the systematic uncertainty we take the product of 
two ${\cal R}$'s with opposite charge configurations of two leptons. 
The methods to repeal the dominant background effects arose from 
two-photon process and particle mis-identification are presented.  
Also, a simulation study is performed to evaluate 
the achievable sensitivity by the $BELLE$ experiment. 
The first goal of this study is to aim for a sensitivity 
of ${\cal O}(10^{-3})$ 
with an integrated luminosity of $10fb^{-1}$. \\
{\ }\\
{\ }\\
\noindent
{\bf 2. ${\cal A}$ vs. ${\cal R}$}\\
\par
In this text we assign the unit vectors of the momenta of positive- and 
negative-charged particles in the final state to 
$\hat{p}_2$ and $\hat{p}_3$, respectively. 
It is obvious that the $T$ and $CP$ reflected states are different. 
Although both transformations change the sign of ${\cal A}$, 
$CP$ reverses the particle's charges while $T$ keeps them unchanged. \\

Let us denote the number of samples with ${\cal A}>0$ and ${\cal A}<0$ 
as $N(l_2^+l_3^-;>)$ and $N(l_2^+l_3^-;<)$, respectively, 
where $l_2$ and $l_3$ are either an electron or a muon of the final 
states, and ${\cal A}>0$ and $<0$ are denoted simply as $>$ and $<$, 
and form the following four ratios, $\cal R$'s, as violation parameters 
instead of the asymmetry, $<{\cal A}>$:
\begin{eqnarray}
{\cal R}^T_{\mu^+ e^-}&\equiv&\frac{N(\mu^+ e^-;>)}{N(\mu^+ e^-;<)}
=\frac{N_o(1+\delta^T_{\mu e})}
{N_o(1-\delta^T_{\mu e})}
=1+2\delta^T_{\mu e}, \\
{\cal R}^T_{e^+ \mu^-}&\equiv&\frac{N(e^+ \mu^-;>)}{N(e^+ \mu^-;<)}
=1+2\delta^T_{e \mu},\\
{\cal R}^{CP}_{\mu^+ e^-}&\equiv&\frac{N(\mu^+ e^-;>)}{N(e^+ \mu^-;<)}
=\frac{N_o(1+\delta^{CP}_{\mu e})}
{N_o(1-\delta^{CP}_{\mu e})}
=1+2\delta^{CP}_{\mu e}, \label{eqn:key4}\\
{\cal R}^{CP}_{e^+ \mu^-}&\equiv&\frac{N(e^+ \mu^-;>)}{N(\mu^+ e^-;<)}
=1+2\delta^{CP}_{e \mu},\label{eqn:key5}
\end{eqnarray}
where the total number of $\mu^+ e^-$ samples, 
$N(\mu^+ e^-;>)+N(\mu^+ e^-;<)$, is 
normalized to be equal to those of the 
$e^+ \mu^-$ samples, $N(e^+ \mu^-;>)+N(e^+ \mu^-;<)$, 
and is expressed as $2 N_o$. 
$\delta^T_{l_2 l_3}$ and $\delta^{CP}_{l_2 l_3}$ 
are the portions of $T$ and $CP$ violations, respectively, 
with a subscript of $l_2 l_3$. \\

If the $CPT$ invariance does not hold, 
the $\delta^{T/CP}_{l_2 l_3}$'s should have different 
non-zero values. 
With $\delta$ and $\Delta$ being $T$ and $CPT$-violating portions 
in the number of samples, respectively, as can be seen in Fig.1, 
the above ${\cal R}$'s can be expressed as 
\begin{eqnarray}
{\cal R}^T_{\mu^+ e^-}&=&{\cal R}^T_{e^+ \mu^-}=1+2\delta, \\
{\cal R}^{CP}_{\mu^+ e^-}&=&1+2(\delta+\Delta); ~~~
{\cal R}^{CP}_{e^+ \mu^-}=1+2(\delta-\Delta), 
\end{eqnarray}
and the $CPT$ violation parameters are also formed as 
\begin{eqnarray}
{\cal R}^{CPT}_{\mu^+ e^-;>}&\equiv&
\frac{N(\mu^+ e^-;>)}{N(e^+ \mu^-;>)} \nonumber\\
={\cal R}^{CPT}_{\mu^+ e^-<0}&\equiv&
\frac{N(\mu^+ e^-;<)}{N(e^+ \mu^-;<)} \nonumber\\
&=&1+2\Delta.
\end{eqnarray}
The $CP$-violation parameters, ${\cal R}^{CP}_{\mu^+ e^-}$ and 
${\cal R}^{CP}_{e^+ \mu^-}$, are obviously not equal. 
Therefore, we can in principle test the $T$, $CP$ and $CPT$ 
invariances by examining the above ${\cal R}$ ratios. 
When $CPT$ holds, but $T$/$CP$ is violated, all four 
$\delta^{T/CP}_{l_2 l_3}$'s have the same non-vanishing value, 
but the ${\cal R}^{CPT}$'s are unity. 
In the following, we assume that $CPT$ invariance holds. \\

\par
The systematic uncertainty is controlled with high precision by 
forming the following $\tilde{\cal R}$, a product of two 
${\cal R}^{T/CP}$'s:
\begin{eqnarray}
\tilde{\cal R} &\equiv& {\cal R}^T_{\mu^+ e^-}\cdot {\cal R}^T_{e^+ \mu^-} 
= {\cal R}^{CP}_{\mu^+ e^-}\cdot {\cal R}^{CP}_{e^+ \mu^-} \nonumber \\
&=&1+4\delta. \label{eqn:key9}
\end{eqnarray}
By factorizing the whole detection efficiency, $\eta$, comprising the 
geometrical acceptances, detection and reconstraction efficiencies, 
as a product of the efficiency, $\eta_1$, independent of 
lepton-charge configuration and a factor $\eta_2$, reflecting 
the efficiency difference due to different charge configurations: 
$\eta(l_2^+ l_3^-;^>_<)\equiv\eta_1(l_2 l_{3};^{>}_{<}~)\cdot 
\eta_2(l_2^+ l_3^-)$, the $\tilde{\cal R}$ is written as 
\begin{eqnarray}
\tilde{\cal R} &=& \frac{N(\mu^+ e^-;>)}{N(\mu^+ e^-;<)}\times
\frac{N(e^+ \mu^-;>)}{N(e^+ \mu^-;<)}\nonumber\\
&=& \frac
{[N_{orig}^{\mu^+ e^-}(1+\delta)\eta_1(\mu e;>)\eta_2(\mu^+e^-)]}
{[N_{orig}^{\mu^+ e^-}(1-\delta)\eta_1(\mu e;<)\eta_2(\mu^+e^-)]} 
\frac
{[N_{orig}^{e^+ \mu^-}(1+\delta)\eta_1(e\mu;>)\eta_2(e^+\mu^-)]}
{[N_{orig}^{e^+ \mu^-}(1-\delta)\eta_1(e\mu;<)\eta_2(e^+\mu^-)]},
\label{eqn:key10}
\end{eqnarray}
where $N_{orig}^{l_2^+ l_3^-}$ is the original number of 
samples produced. 
Since the sample of $\mu^{\pm} e^{\mp}>0$ has the same 
geometrical configuration, but opposite charge configuration, 
to the sample of $e^{\pm} \mu^{\mp}<0$, 
$\eta_1(\mu e;^{>}_{<})\cong \eta_2(e \mu;^{<}_{>}~0)$ could be 
valid. 
On the other hand, the $\eta_2(\mu^{\pm} e^{\mp})$'s cancel each 
other out in the same reaction. 
Therefore, a high degree of accurate cancellation is expected 
in the form of $\tilde{\cal R}$. 
The normalization between the charge-conjugated reactions used in 
eqs.(\ref{eqn:key4}) and (\ref{eqn:key5}) is not necessary 
in this case. 
Also, long-term instabilities of the experimental situations, 
such as the beam conditions and detector performances, 
do not affect $\tilde{\cal R}$. \\
{\ }\\
{\ }\\
\noindent
{\bf 3. Sensitivity vs. Background}\\
\par
Including the background, $N_{BG}$, the $\tilde{\cal R}$ ratio 
is expressed as 
\begin{equation}
\tilde{{\cal R}} = ~1+4\delta~+~
\{\varepsilon(\mu^+ e^-;>)-\varepsilon(\mu^+ e^-;<)+
\varepsilon(e^+ \mu^-;>)-\varepsilon(e^+ \mu^-;<)\},
\end{equation}
where the third term is the background contributions, and 
each $\varepsilon$ is the ratio of $N_{BG}$ corresponding to 
the signal samples. 
The statistical sensitivity is approximated as 
\begin{equation}
(\frac{\Delta \tilde{\cal R}}{\tilde{\cal R}})^2 = 
4~[ (\frac{\Delta N_o}{N_o})^2 + (\frac{\Delta N_{BG}}{N_o})^2 ],
\label{eqn:key12}
\end{equation}
where $\Delta N_{BG}$ is the uncertainty of $N_{BG}$, and 
is assumed to be the same in the four different kinds of samples. 
The background does not affect the statistical sensitivity 
as long as $\Delta N_{BG}\ll\Delta N_o$ is satisfied: 
$\Delta \delta=1/(2 \sqrt{N_o})$. 
(Hereafter, the subscript of $l_2 l_3$ of the single ${\cal R}$ 
parameter, ${\cal R}^T$ and ${\cal R}^{CP}$, is omitted 
unless it is essential.) \\ 

The dominant physics background arises from a two-photon process, 
especially, $e^+e^-\rightarrow e^+e^-\mu^+\mu^-$, whose contamination 
could be a few percent of the signal samples. 
Since the distributions of the $\mu^+$ and $\mu^-$ of the process 
is symmetric, the two-photon backgrounds, $N^{\gamma\gamma}$, 
in $N(e^+\mu^-)$ and 
$N(\mu^+ e^-)$ samples can be estimated just by $N(e^+\mu^+)$ and 
$N(\mu^- e^-)$ of two-photon samples, respectively. 
Therefore, $\Delta N^{\gamma\gamma}$ 
is the statistical accuracy of the observed $N(e^{\pm}\mu^{\pm})$ 
samples, so that the second term on the right-hand side 
in eq.(\ref{eqn:key12}) can be ignored compared to the first one, 
as evaluated in the next section. 
Furthermore, since the observed ${\cal R}^T$ ratio is expressed as 
\begin{eqnarray}
{\cal R}^{T;obs}&=&\frac{(N^{obs};>)}{(N^{obs};<)}
=\frac{ (N^{\tau\tau};>) +  (N^{\gamma\gamma};>) }
      { (N^{\tau\tau};<) +  (N^{\gamma\gamma};<) }\nonumber\\
&=&{\cal R}^{T;true} (~ 1 + \zeta_{\gamma\gamma} ~), \\
\zeta_{\gamma\gamma} &\equiv& 
~\alpha_{\gamma\gamma}(1-\frac{{\cal R}^{T;obs}}
{\beta_{\gamma\gamma}^{T;obs}}
\end{eqnarray}
where the signal sample is here denoted as $N^{\tau\tau}$ 
to distinguish it from $N^{\gamma\gamma}$. 
Here, $\alpha_{\gamma\gamma}$
(=$(N^{\gamma\gamma};>)/(N^{\tau\tau};>)$) is 
the contamination rate, and ${\cal R}^{T; true}$, 
${\cal R}^{T; obs}$ and $\beta_{\gamma\gamma}^{T; obs}$ 
are the ${\cal R}^T$ ratios formed by $N^{\tau\tau}$, 
$N^{obs}$ and $N^{\gamma\gamma}$ samples, respectively. 
Any artificial background asymmetry, if it exists, appears 
to reduce its size 
by $\alpha_{\gamma\gamma}$ times, and then becomes ineffectual. 
The statistical uncertainty of $\zeta_{\gamma\gamma}$ is 
$\Delta\zeta_{\gamma\gamma}\approx 
\sqrt{2\alpha_{\gamma\gamma}}/\sqrt{N^{\tau\tau}}$. \\

Another dominant background sample comprises, as will be shown later 
when discussing a simulation study, an electron from 
$\tau\rightarrow e\nu\overline{\nu}$ decay and a pion 
predominantly from $\tau\rightarrow\pi\nu$. 
This is because a small fraction of hadrons would be mis-identified 
as muons by a muon counter due to their punch through, 
while electrons would be correctly identified with high purity 
by an electromagnetic calorimeter. 
With this knowledge in mind, we argue on how to control the effect 
of the mis-identified samples on ${\cal R}^T$. 
Let us first suppose to have prepared an $e+X$ sample set, $N^{obs}_{e+X}$,  
by applying some kinematical conditions and electron identification 
on collected data samples, as is done in the next section. 
The sample comprises an electron and an opposite-charged 
particle, $X$, being a muon or hadron. 
Next, relying upon a muon identification, classify the samples into 
$e+\mu$ samples, $N^{obs}_{\mu}$, or $e+\Pi$ samples, $N^{obs}_{\Pi}$, 
where $\Pi$ means a hadron. 
With $\kappa_{\mu}$ and $\kappa_{\Pi}$ being the probabilities of a muon 
and a hadron to be classified as a muon, respectively, 
and $N^{true}_{\mu}$ and $N^{true}_{\Pi}$ being the true numbers of 
$e+\mu$ and $e+\Pi$ samples in $e+X$, these samples 
relate among them as 
$N^{obs}_{\mu} = \kappa_{\mu} N^{true}_{\mu} + 
        \kappa_{\Pi}N^{true}_{\Pi}$, 
$N^{obs}_{\Pi} = (1-\kappa_{\mu}) N^{true}_{\mu} + 
        (1-\kappa_{\Pi}) N^{true}_{\Pi}$, and 
$N^{obs}_X=N^{obs}_{\mu}+N^{obs}_{\Pi}=N^{true}_\mu+N^{true}_\Pi.$
The observed ${\cal R}^T$-ratio is then expressed as 
\begin{eqnarray}
{\cal R}^{T;obs}&=&\frac{(N^{obs}_{\mu};>)}{(N^{obs}_{\mu};<)}
=\frac{ \kappa_{\mu} (N^{true}_{\mu};>) + 
        \kappa_{\Pi} (N^{true}_{\Pi};>) }
      { \kappa_{\mu} (N^{true}_{\mu};<) + 
        \kappa_{\Pi} (N^{true}_{\Pi};<) }
\nonumber\\
&=&{\cal R}^{T; true} (~ 1 + \zeta_{\Pi} ~), \\
\zeta_{\Pi} &\equiv& \frac{\kappa_{\Pi}}{\kappa_{\mu}}~
[~ \frac{(N^{true}_{\Pi};>)}{(N^{true}_{\mu};>)} -
  \frac{(N^{true}_{\Pi};<)}{(N^{true}_{\mu};<)}~ ],\label{eqn:key17}
\end{eqnarray}
where the ${\cal R}^{T; true}$ is formed by $N_{\mu}^{true}$ samples, and 
$\kappa_{\mu}$ and $\kappa_{\Pi}$ are irrelevant of the sign 
of $\cal A$. 
The $\zeta_{\Pi}$ is approximated in terms of the observed 
ratios as 
\begin{eqnarray}
\zeta_{\Pi} \approx \kappa_{\Pi}\alpha_X(~1-
\frac{{\cal R}^{T; obs}}{\beta_X^{T; obs}}~),\label{eqn:key18}
\end{eqnarray}
where $\alpha_X=(N_X^{obs};>)/(N_{\mu}^{obs};>)$ 
and $\beta_X^{T; obs}$ is a ${\cal R}^{T; obs}$ ratio 
formed by the $e+X$ samples. 
Since $\kappa_{\mu}$ and $\kappa_{\Pi}$ are in the most cases about 90\% 
and 1\%, respectively, any higher order term in an approximation of 
eq.(\ref{eqn:key18}) can be disregarded.  
\noindent
The $\zeta_{\Pi}$ is thus a product of two elements. 
One is the detector performance of the hadron mis-identification 
probability, and the other is $\alpha_X$ times 
the ${\cal R}^{T; obs}$ and $\beta^{T; obs}_X$ difference. 
The magnitude of the former element is on an order of 
${\cal O}(10^{-2})$, and the latter would be zero within the statistical 
accuracy under $T/CP$ invariance. 
By taking into account the correlation between ${\cal R}^{T; obs}$ and 
$\beta^{T; obs}_X$, the statistical uncertainty of $\zeta_{\Pi}$ 
is $\Delta\zeta_{\Pi}\approx \kappa_{\Pi}
\sqrt{2\alpha_X(1+\alpha_X)}/\sqrt{N_{\mu}^{obs}}$. \\

\par
It is worth mentioning that since both statistical uncertainties of 
$\zeta_{\gamma\gamma}$ and $\zeta_{\Pi}$ are inversely
proportional to the squareroot of the signal samples, 
the achievable sensitivity on $\delta$ is 
improved with increased integrated luminosity. \\
{\ }\\
{\ }\\
\noindent
{\bf 4. Simulation Study}\\
\par
Using a fast $BELLE$ detector-simulator [8] for $p_T>$0.1 GeV/c tracks, 
the $QQ$ generator [9] and $KORALB$ [10] are used for 
$e^+e^-\mu^+\mu^-$ two-photon process, $B\overline{B}$, 
continuum, and lepton-pair productions, and $\tau$-pair production, 
respectively, in the collision of a 3.5-GeV positron and 
an 8-GeV electron for an integrated luminosity of 10$fb^{-1}$. 
The following selection criteria are imposed on generated samples, 
where kinematical variables are evaluated in the lab-frame:
\begin{enumerate}
\item   The sample should comprise only 2 charged tracks 
        with opposite charges and no photon. 
        The photon is defined as having an energy of $E>$100 MeV  
        (cut-1).
\item   In order to remove lepton-pair productions, 
        four conditions are required: 
        Each track should have a momentum of $p<$6 GeV/c; 
        their sum should be $p_{sum}<$9 GeV/c; 
        the missing transverse momentum squared would be 
        $(\Sigma p_t)^2 >$0.02 (GeV/c)$^2$; 
        the direction of the missing momentum should be 
        $\cos\theta^{missing}<$0.990 (cut-2). 
\item   The polar-angle and momentum apartures for tracks are set to fit 
        those of both an electromagnetic calorimeter ($ECL$) and 
        a muon counter ($KLM$: $\kappa_{\mu}=90\%$ and $\kappa_{\Pi}=1\%$) 
        of the barrel region 
        as -0.642$<\cos\theta<$0.860 and $P>$0.5 GeV/c (cut-3).
\item   A track which meets the following conditions is 
        assigned as an electron: 
        $ECL$ counts it as an electron, but $KLM$ does not count it 
        as a muon, 
        and the electron probability evaluated by a set of the central 
        tracking chamber ($CDC$), the aerogel Cerenkov counter
        ($AER$), and the time-of-flight counter ($TOF$) is $>$1\%. \\
        A track with $p>$1.2 GeV/c is assigned as $X$ 
        when it is not regarded as an electron by $ECL$, and 
        its evaluated probability by the $CDC/AER/TOF$ to be either 
        a muon or any hadron is $>$1\%. 
        The requirement $p>$1.2 GeV/c is imposed due to the $KLM$ 
        performance for particle identification. \\
        The classification of $X$ into $\mu$ or $\Pi$ is performed 
        according to $KLM$ information. 
\end{enumerate}
\noindent
The resulting samples which pass the above criteria are summarized 
in Table 1. \\

Concerning the two-photon process, the equivalent photon approximation 
$QQ$ generator employed might not be accurate enough in the case 
an electron or a positron scattered at a large angle. 
Therefore, their numbers given in Table 1 should be considered 
to be a brief estimate. 
However, the arguments discussed in the previous section are 
indeed valid. 
According to Table 1, $N^{\tau\tau}$ (=$N_o$) =34.8K and 
$N^{\gamma\gamma}\approx$ 700: 
$\alpha_{\gamma\gamma}$ is $2\%$ and 
the second term of eq.(\ref{eqn:key12}) is $1/50$-times smaller 
than the first one, and thus can be disregarded. 
$\Delta\zeta_{\gamma\gamma}$ is $\approx 1\times 10^{-3}$. 
The two-photon $e^+e^-\mu^+\mu^-$ background thus does not yield 
an appreciable contribution. \\

\begin{table}[b]
\footnotesize
\begin{description}
\item[{Table 1:}] Selected sample rates satisfying the selection criteria 
discussed in the text. $B\overline{B}$, continuum, and lepton pair 
samples are generated for an integrated luminosity of 10$fb^{-1}$. 
For $e\overline{e}\mu\overline{\mu}$ two-photon reaction, 
the resulting numbers for a generated 10M samples are normalized to 700M. 
\end{description}
\begin{center}
\begin{tabular}{lrrrrrrr}
\hline\hline
\noalign{\vspace{3pt}} 
 Mode & ~$B^o\overline{B}^o$ & $B^+B^-$ & conti. &
~~~$\tau\overline{\tau}$ & ~~~$\mu\overline{\mu}$ & 
~~~$e\overline{e}$ & $e\overline{e}\mu\overline{\mu}$ \\
\noalign{\vspace{3pt}} 
\hline
\noalign{\vspace{3pt}} 
 Generated samples & 7.5M & 7.5M & 45M & 13.5M & 13.5M & 13.5M & 700M \\
\hline
 Selected samples \\
 $e^+\mu^-$ and $e^-\mu^+$ & 3 & 0 & 5 & 139.3K & 0 & 0 & 2.8K\\
 $e^+ e^-$ & 2 & 0 & 4 & 78.3K & 0 & 434 & 0 \\
 $\mu^+ \mu^-$ & 2 & 1 & 5 & 60.1K & 32.1K & 0 & 70.5K \\
\noalign{\vspace{3pt}}
\hline\hline
\end{tabular}
\end{center}
\end{table}
\small

Since $B\overline{B}$ and continuum samples produce high
multiplicities of both charged tracks and photons, 
they are reduced to a negligible level.  
Although $\mu$-pair samples cannot be kinematically removed by cut-2, 
their contaminations in the $e\mu$ samples are sufficiently rejected 
by the particle-identification performance. \\

For $\tau$-pair production, the selected and classified sample rates 
are listed in Table 2. 
The acceptance for the $e\mu$ reaction is 1\% to the generated 
$\tau$-pair samples, and the expected total number of samples is 
$N_{e\mu}$=139K, among which mis-identified samples amount to 1\%. 
The $\alpha_X$ is about 2 in this case, and the $\Delta\zeta_{\Pi}$ is 
$\approx 2 \times 10^{-4}$. 
The expected statistical sensitivity is $\Delta\tilde{\cal R}$=0.01 
or $\delta$=0.0027. 
The ${\cal A}$ asymmetry parameters calculated 
in the CM-frame are given in Fig.2. 
The dip structure at ${\cal A}$=0 is due to an effect of 
the $(\Sigma p_t)^2 >$0.02 (GeV/c)$^2$ and $\cos\theta^{missing}$ cuts. 
Removing these cuts yields a 10\% increase of the $N_o$ samples, 
and makes $N^{\gamma\gamma}$ samples twice but enhances 
$\Delta\zeta_{\gamma\gamma}$ only by $\sqrt{2}$ times. \\

Among the $e\mu$ samples, the backgrounds whose tracks are correctly 
particle-identified comprise a muon from 
$\tau\rightarrow\mu\nu\overline{\nu}$ and an electron from 
a single Dalitz mode of $\pi^o$ decay following the decay sequence 
of $\tau\rightarrow\rho\nu$ and $\rho\rightarrow\pi^{\pm}\pi^o$. 
The rate of this background consists of only 9 samples for 13.5M 
generated $\tau$-pairs. 
We can therefore disregard it. \\

The above fact causes us to consider another background. 
Namely, an electron produced through pair conversion or Compton 
scattering of photons by detector materials could play a similar 
role as the electron from single Dalitz decay. 
In order to briefly evaluate this background, we calculated the rate 
of (single $\mu$-track) + (single photon) samples for which the momentum 
and polar angle apartures are set to be the same as those of cut-3. 
It is 4$\times$10$^{-4}$ to the generated $\tau$-pairs, or 7\% to 
the final $e\mu$ samples. 
The material thickness around the collision point is 
2\% of the radiation length for the $BELLE$ detector [7,8]. 
This yields $\approx$2\% of the pair conversion and $\approx$0.2\% of 
the Compton-scattering rates for photons with $E>$500MeV. 
Presuming that the probability for only one of the pair-conversion 
electrons detected is less than 10\%, a photon would yield 
an objective electron by 0.5\% at most. 
As a result, the background rate to $e\mu$ samples could be 
3$\times$10$^{-4}$ or less, so that its effect on 
$\Delta\tilde{\cal R}$ can be totally disregarded. \\

Consequently, with an integrated luminosity of 10$fb^{-1}$, 
we can reach a $\delta$ sensitivity of 0.003, for which 
the background effects are at most 1-2 orders of magnitude small. 
Furthermore, a 100$fb^{-1}$ accumulated luminosity allows us 
to achieve a sensitivity of 0.001. \\
\begin{table}[t]
\footnotesize
\begin{description}
\item[{Table 2:}] Selected and classified sample rates for 13.5M $\tau$-pairs. 
The right-most column in (b) indicates the contamination rate 
due to particle mis-identification. 
\end{description}
\begin{center}
\begin{tabular}{lrrr}
\hline\hline
\noalign{\vspace{3pt}} 
(a)& Samples passed cut-1: &    2,472K & \hspace*{5.8cm} \\
   &                cut-2: &    2,051K \\
   &                cut-3: &    1,124K \\
\noalign{\vspace{3pt}} 
\end{tabular}
%
\begin{tabular}{lrcrrr}
\hline
\noalign{\vspace{3pt}} 
(b)& Classified mode:& Samples & Accepted rates(\%)& 
mis-$PID$'ed rates(\%) \\
\noalign{\vspace{2pt}} 
 &  $e^{\pm}\mu^{\mp}$:&  139.3K  & 2$\times$0.52\% & 1.1\% \\
 &  $e^+e^-$:&   78.3K  &          0.58\% & 0.04\% \\     
 &  $\mu^+\mu^-$:&   60.1K  &          0.45\% & 2.1\% \\
\noalign{\vspace{2pt}} 
\hline
\end{tabular}
%
\begin{tabular}{lllrrrr}
\noalign{\vspace{2pt}} 
 & & mode: & ~~~$N_{e X}$ & ~~~$N_{\mu}(\%)$ & ~~~$N_{\Pi}(\%)$ & 
\hspace*{2.3 cm} \\
(c)& Contents of $N_X$ & $N_{e^+ X}$: & 145.1K & 48.2 & 51.8  \\
 & & $N_{e^- X}$: & 144.5K & 48.0 & 52.0  \\
\noalign{\vspace{2pt}} 
\hline\hline
\end{tabular}
\end{center}
\end{table}
{\ }\\
{\ }\\
\noindent
{\bf 5. Conclusion}\\
\par
In addition to studying the $CP$ violation of $B$ mesons, the $B$-Factory 
would also provide a good playground for testing the $T/CP$ 
invariance of the $\tau$ lepton in the reaction of $e^+ e^-\rightarrow
\tau^+\tau^-$ with $\tau\rightarrow \mu/e\nu\overline{\nu}$. 
This experiment is rather simple and does not require any 
spin-related information, such as a polarized beam or a polarization 
measurement of the final state leptons. 
With a 10$fb^{-1}$ of integrated luminosity, the $T/CP$ invariance 
can be tested with an accuracy of ${\cal O}(10^{-3})$. 
Furthermore, both the statistic and systematic sensitivities could be 
improved with additional data samples. \\

Other $\tilde{\cal R}$ ratios, such as ${\cal R}_{e^+e^-}$, 
${\cal R}_{\mu^+\mu^-}$, $\tilde{\cal R}_{e\pi}$, $\tilde{\cal R}_{\mu\pi}$,
and $\tilde{\cal R}_{\pi\pi}$, can also be constructed with 
a $\pi$ from $\tau\rightarrow\pi\nu$ decay, in addition to a muon 
and an electron from $\tau\rightarrow\mu/e\nu\overline{\nu}$. 
For those cases, contamination of 
$\mu^+\mu^-$ pair production should be brought down to a sufficiently 
small level; also, high purity of pion identification is indispensable. \\
{\ }\\
{\ }\\
\noindent
{\bf Acknowledgement}\\
\par
We gratefully acknowledge Professors S. Uehara and S. Kitakado for their 
advice concerning the two-photon process and 
Professors H. Ozaki, A. I. Sanda, Y. Okada, M. Tanimoto, and 
M. Tanaka for many useful discussions. 
This work was partly supported by Grant-in-Aid for Scientific 
Research on Priority Areas (Physics of $CP$ violation) from 
the Ministry of Education, Science, and Culture of Japan. \\

\newpage
{\ }\\
\noindent
{\bf References}
\begin{description}
\small
\item[{[1]}] For instance, see {\em Symmetries and Fundamental 
        Interactions in Nuclei}, ed. by W. C. Haxton and E. H. Henley, 
        World Scientific, 1995. 
\item[{[2]}] H. Burkard et al., Phys. Lett. 160B (1985) 343.
\item[{[3]}] T. D. Lee, 1994 International Workshop on $B~ PHYSICS$, 
        Nagoya, Japan, 26-28 October 1994, eds. by A. I. Sanda and S. Suzuki, 
        World Scientific. 
\item[{[4]}] S. Weinberg, Phys. Rev. Lett., 37 (1976) 657.
\item[{[5]}] See, for an example, H. P. Nilles, Phys. Rep., 
        C110 (1984) 1.
\item[{[6]}] See, for an example, M. Chaichian and K. Huitu,
        Phys. Lett., B384 (1996) 157; 
        J-H. Kim, J. K. Lee and J. Sik, Phys. Rev. D55 (1997) 7296.
\item[{[7]}] For the description on $KEK-B$ Factory and $BELLE$
        detector, for instance, see J. Haba, Nucl. Inst. and Meth., 
        A368 (1995) 74; T. Nozaki, Nucl. Phys. B(Proc. Suppl.) 50 
        (1996) 288, and also ref.[8].
\item[{[8]}] M. Aoki et. al, Nucl. Instr. and Meth. A366 (1995) 85;
        Letter of Intent, The $BELLE$ collaboration, KEK Report 
        94-2 (1994).
\item[{[9]}] P. C. Kim, ``MC NEWS'', CLEO internal report, 
        unpublished.
\item[{[10]}] S. Jadach and Z. Was, Comput. Phys. Commun. 85 (1995) 
        453. 
\end{description}
\normalsize
\newpage
{\ }\\
{\ }\\
\noindent
{\bf Figure Captions}
\small
\begin{description}
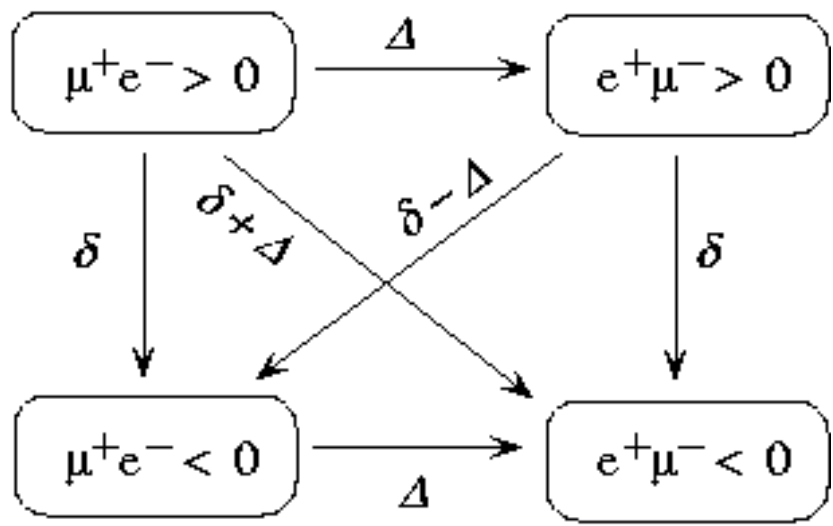
\item[{Fig.1:}] Schematic relation among $T$, $CP$, and $CPT$
        reflections. 
        $l_2^+ l_3^-~^>_<0$ denotes the reactions which the positive 
        and negative charged particles, $l_2$ and $l_3$, form 
        either ${\cal A}>0$ or ${\cal A}<0$. 
        The $T$ transformation changes the sign of ${\cal A}$, and 
        $\delta$ denotes its violation portion in the data samples. 
        $CP$ changes the particles to anti-particles and 
        reverses the sign of ${\cal A}$ according to our convention. 
        $CPT$, whose violation portion is denoted as $\Delta$, 
        also changes the particles to anti-particles, but not 
        the sign of ${\cal A}$. 
        Any triangle route results in a vanishing violation, 
        as it must. 
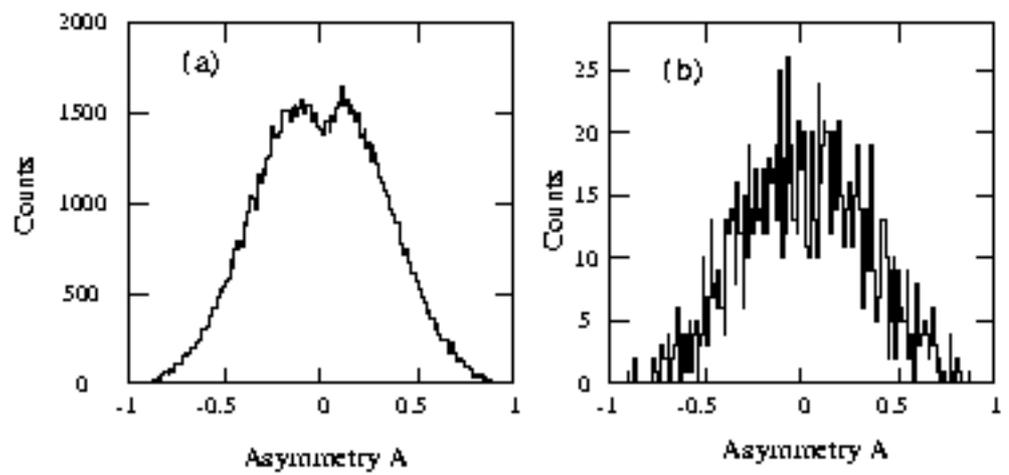
\item[{Fig.2:}] Resulting $\cal A$ distributions for 13.5 M 
        $\tau$-pairs, corresponding to an integrated luminosity 
        of 10$fb^{-1}$. (a) is for the samples selected as 
        $\mu e$ according to the criteria mentioned in the text. 
        (b) is for the particle mis-identified background contamination 
        in the sample of (a). 
\end{description}
\normalsize
\end{document}